\documentclass[preprint, 3p, twocolumn]{elsarticle}
\usepackage{graphicx}
\usepackage{booktabs}
\usepackage{tabularx}
\usepackage[normalem]{ulem}
 \useunder{\uline}{\ul}{}
 \usepackage{rotating}
 \usepackage{dsfont}
 \usepackage{textcomp}
\usepackage{gensymb}
\usepackage{color}
\def\Manu#1{\textcolor{black}{#1}}
\newcommand{\bit}[1]{\textit{\textbf{#1}}}
\usepackage[switch]{lineno}
\usepackage{graphicx,xcolor}
\usepackage{epstopdf}
\usepackage{setspace}
\usepackage[T1]{fontenc}
\usepackage{amssymb,amsfonts,amsmath}
\usepackage{setspace}
\usepackage{subcaption}
\usepackage{mathtools}
\usepackage{etoolbox}
\biboptions{sort&compress}
\newcolumntype{Y}{>{\centering\arraybackslash}X}

\begin{document}

\title{\Manu{Bubble coalescence in worm-like micellar solutions}}
\author[vinny,equal]{V. Chandran Suja}
\ead{vinny@stanford.com}
\author[vinny,equal]{A. Kannan}
\author[bruce]{B. Kubicka}
\author[alex]{A. Hadidi}
\author[vinny]{G.G. Fuller}\ead{ggf@stanford.com}
\address[vinny]{Department of Chemical Engineering, Stanford University, Stanford, California 94305}
\address[alex]{Department of Mechanical Engineering, UCLA,  Los Angeles, CA 90095}
\address[bruce]{Department of Mechanical Engineering, Cornell, Ithaca, NY 14850}
\address[equal]{Authors contributed equally}

\begin{abstract}
Surfactants in aqueous solutions self-assemble in the presence of salt, to form long, flexible worm-like micelles (WLM). WLM solutions exhibit viscoelastic properties and are used in many applications, such as for cosmetic products, drag reduction and hydraulic fracturing. The dynamics of bubbles in WLM solutions are important considerations for the stability of many of these products. In this manuscript, we investigate the thin film drainage dynamics leading up to the coalescence of bubbles in WLM solutions. The salts and surfactant type and concentrations were chosen so as to have the viscoelastic properties of the tested WLM solutions span over two orders of magnitude in moduli and relaxation times. The various stages in drainage and coalescence - the formation of a thick region at the apex - a dimple, the thinning and washout of this dimple and the final stages of drainage before rupture, are modified by the viscoelasticity of the wormlike micellar solutions. As a result of the unique viscoelastic properties of the WLM solutions, we also observe a number of interesting fluid dynamic phenomena during the drainage process including elastic recoil, thin film ripping and single-step terminal drainage.       
\end{abstract}

\maketitle

\section{Introduction}

Micelles are self-assembled surfactant molecules comprising of a hydrophilic head and a hydrophobic tail. In a micelle formed in an aqueous solution, the hydrophobic group is hidden within the core, exposing the hydrophilic heads. Most surfactants self-assemble in solution above their critical micellar concentration (CMC), to form spherical micelles. In some cases, particularly in the presence of salt, these molecules form long, flexible chains, which are called worm-like micelles (WLM) \cite{belmonte2000self,cates1990statics}. At a high enough concentration, these supramolecular chains interact and entangle in the same manner as long chained polymers \cite{wheeler1996structure,walker2001rheology}. This gives the solution its unique viscoelastic properties. 

\begin{figure*}[!th]
\centering
\includegraphics[width=\linewidth]{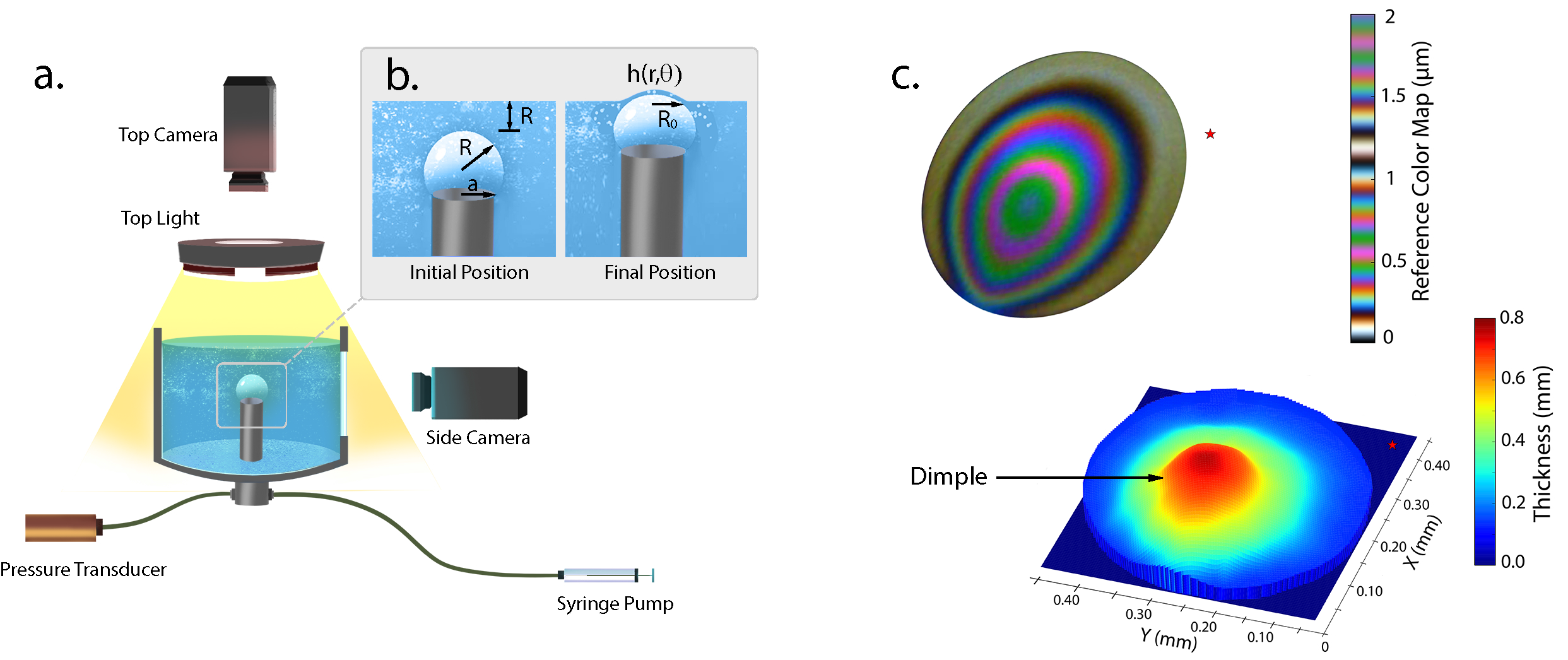}
\caption{Schematic of the Dynamic Fluid-Film Interferometer (DFI) used to perform the reported experiments and a typical interferogram obtained from the experiments with worm-like micelles. ({\bf a.}) The setup with the labeled components. ({\bf b.}) The initial and final positions of the bubble during an experiment. Here $R$ is the radius of curvature of the bubble, $h(r,\theta)$ is the film thickness as a function of the radial position ($r$) and angular position $\theta$, and $R_0$ is the radial extent of the film visible on the interferogram. ({\bf c.}) An interferogram showing the asymmetric dissipation of a dimple (the thick region in the film) following its instability. The adjacent tile  shows physical film thickness corresponding to the interferogram. A star  is used to mark the corresponding points in the interferogram and the reconstruction. }\label{fig:ExperimentalSetup}
\end{figure*} 

The rheology of the worm-like micelles are intrinsically connected to the morphology of the micelles, which change with the surfactant concentration ($C_D$) and the salt concentration ($C_S$). The degree of entanglement of these micelles change as a function of the salt to surfactant ratio $(C_SC_D^{-1})$ up to a critical value close to one, with lower $C_SC_D^{-1}$ behaving like low molecular weight polymer solutions and at higher values like high molecular weight polymer solutions \cite{shikata1987micelle}. In the low $C_SC_D^{-1}$ limit, their behavior can be described by the Rouse and Zimm theories. Further, it was observed that as $C_SC_D^{-1}$ increased, the length of the chains increased, without any noticeable change in the diameter of the micelles \cite{shikata1987enormously}. Above a critical value of $C_SC_D^{-1}$, their behavior can be modeled by a Maxwell model with a constant plateau modulus ($E_0$) and a single relaxation time ($\lambda$); with $E_0$ proportional to $C_D^{-2.2}$ and independent of $C_SC_D^{-1}$, and $\lambda$ decreasing with increasing $C_SC_D^{-1}$ \cite{shikata1987micelle}. This is a consequence of stress relaxation through chain breakage becoming energetically favorable over reptation ($\tau_b \ll \tau_{rep}$) at higher $C_SC_D^{-1}$, resulting in a behavior similar to a Maxwell fluid with a relaxation time of $\tau_m=\sqrt{\tau_b\tau_{rep}}$ \cite{dreiss2017wormlike,cates1990statics, wheeler1996structure}.

Worm-like micelles are commonly used in many applications \cite{dreiss2017wormlike,yang2002viscoelastic} due their interesting rheological properties, both in the linear regime \cite{berret1993linear,turner1991linear,walker2001rheology,shikata1987micelle}
as well as the non-linear regime \cite{spenley1993nonlinear,khatory1993linear}. Noteworthy applications include the use of WLM solutions as bases for cosmetic products, drag reduction agents and hydraulic fracturing agents. In many of these applications, WLM solutions encounter both solid particles and gas bubbles. As such, a number of studies have been focused on documenting the dynamics of particles and bubbles in viscoelastic WLM solutions.  Unlike in purely viscous liquids, studies have revealed that a settling solid particle never reaches a terminal velocity, and instead exhibits oscillatory motion in the stream wise direction \cite{mohammadigoushki2016sedimentation,kumar2012oscillatory,akers2006impact,jayaraman2003oscillations}. Others studies have also documented the impact of aqueous drops on WLM films  \cite{lampe2005impact}, the impact of WLM drops on hydrophobic substrates \cite{cooper2002drop} and the pinch-off dynamics of WLM drops \cite{smolka2003drop}.

The dynamics of bubbles in WLM solutions has also garnered significant attention. Similar to the solid particle case, the rise of bubbles in a worm-like micellar solution is known to exhibit oscillatory behavior due to the non-Newtonian nature of the fluid \cite{belmonte2000self,handzy2004oscillatory}. The elasticity of WLM solutions also leads to interesting trends during the rupture dynamics of bubbles \cite{tammaro2018elasticity,sabadini2014elasticity}. Despite the rich literature surrounding bubble interactions in WLM solutions, the important events leading up to bubble coalescence in WLM solutions has not been explored.

\begin{figure*}[!h]
\centering
\includegraphics[width=\linewidth]{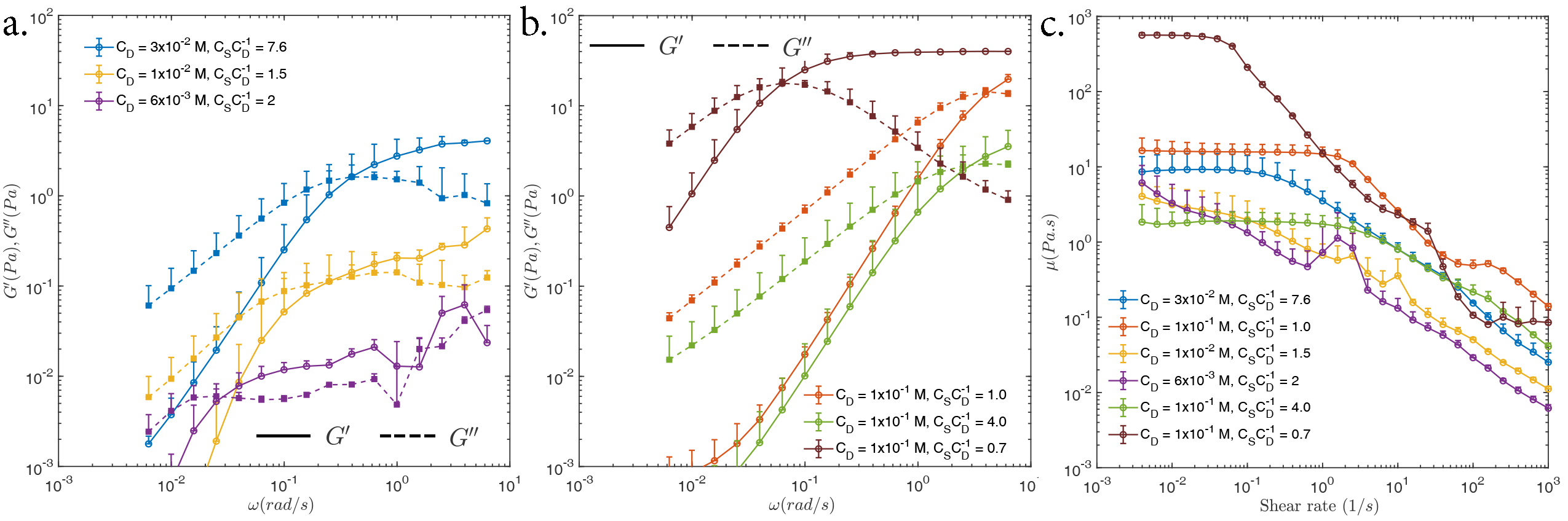}
\caption{Dynamic oscillatory shear rheology and viscosity measurements of the tested worm-like micellar solutions. {\bf a} The elastic ($G'$) and the viscous ($G''$) modulus of solutions with low surfactant concentration ($C_D<0.01\; mol\;L^{-1}$) and different salt concentrations. The peak elastic and viscous moduli vary over two orders of magnitude across the samples. {\bf b} The elastic ($G'$) and viscous ($G''$) modulus of solutions with surfactant concentration $C_D = 0.1\; mol\;L^{-1}$ and different salt concentrations. The peak elastic and viscous moduli are comparable across the samples, while the cross-over frequency varies by two orders of magnitude.  {\bf c} Viscosity measurements of all tested samples. Note: the one sided error bars are standard deviations measured across three separate measurements. See \bit{Supplementary Information Table} for the numerical values of the relaxation time, zero shear viscosity and plateau modulus.}
\label{fig:rheology}
\end{figure*}

In many of the applications of WLM fluids such as cosmetic products, the stability of air bubbles against coalescence is an important quality control. The drainage of the fluid trapped between these air-solution interfaces and the subsequent coalescence process can be studied with the help of the model system of a single air bubble coalescing against a flat air-solution interface \cite{frostad2016dynamic}. A custom built instrument, referred to as the dynamic fluid-film interferometer (DFI) has been used in the past to study the bubble drainage in non-aqueous lubricants \cite{suja2018evaporation,suja2020foam} and aqueous surfactant \cite{frostad2016dynamic} and protein \cite{kannan2018monoclonal,kannan2019linking} solutions. 

In this work, we used the DFI to study the coalescence of bubbles in worm-like micellar solutions consisting of the surfactant, hexadecyltrimethylammonium bromide (CTAB) and the salt, sodium salicylate (NaSal) at different surfactant and salt concentrations. In section \ref{sec:methods}, we detail the characterization techniques and experimental protocols. In section \ref{sec:results}, we report the drainage dynamics of bubbles in WLM solutions, focusing both at the onset of capillary drainage and the final stages of film thinning before coalescence. These results link the phenomena of thin film drainage with the relaxation processes observed in these self-assembled structures, and thus advance the understanding of bubble stability in WLM fluids. 

\section{Materials and Methods}\label{sec:methods}

\subsection{Sample Preparation}\label{sec:sample}
Solutions with worm-like micelles were prepared by dissolving a cationic surfactant, hexadecyltrimethylammonium bromide (CTAB) and a salt, sodium salicylate (NaSal) in water at various concentrations ($C_D$) and ($C_S$) respectively to span different regimes in viscoelasticity. During sample preparation, the surfactant was dissolved first and the salt was added slowly with stirring, to avoid clumping and heterogeneity. Subsequently, the samples were equilibrated for at least two days before performing the experiments. 

\subsection{Bulk Rheology}

Shear rheology was measured using the TA Instruments Discovery HR-3 rheometer with a $2\degree$ cone and plate geometry (20 mm radius). Before each experiment, surfaces were cleaned with ethanol and water. $0.625$ ml of the WLM solution was added to the plate and the cone was lowered to fill the gap of $50\; \mu m$ with the sample. Frequency sweeps were performed in the range $6.28 \times 10^{-3}$ to $6.28$ rad/s ($10^{-3}$ to $1$ $Hz$)  to determine the elastic and viscous moduli for the solutions.

\subsubsection{Experimental Setup}

The thin film drainage experiments reported in the manuscript was performed using the so called dynamic fluid-film interferometer, hereon DFI (Fig.\ \ref{fig:ExperimentalSetup}a). The specific details regarding its construction is available in Frostad et al. \cite{frostad2016dynamic} and in references therein.  Prior to the start of every experiment reported here, the selected WLM solution was added to a Delrin chamber, and an air bubble of radius $R$ and volume $1.2 \pm 0.15 \mu L$ was created through a 16-gauge blunt tipped needle (1.194 mm ID; 1.651 mm OD). This bubble volume was selected to avoid any instabilities associated with manipulating bubbles on capillaries \cite{chandran2016impact}. The bubble was formed using a syringe pump (Harvard Apparatus Pico
Plus 11, Harvard Apparatus, Holliston, MA) and the pressure was measured using a transducer (Omega PX409, OMEGA Engineering Inc., Stamford, CT).  A camera (ThorLabs DCU223C) imaged the bubble from the side, providing real time feed back on the bubble position and volume. A DC Servo Actuator (Newport TRB25CC) provided vertical translation and lowered the chamber to position the bubble at a distance $R$ from the flat air-solution interface (Fig.\ \ref{fig:ExperimentalSetup}b).  This was the initial state of the system before all the experiments.

At this point the experiment starts with the pressure transducer measuring the pressure inside the bubble at $20\;Hz$.  After the pressure was monitored for 10 seconds (to make sure there are no fluctuations in the size of the bubble), the air-solution interface was lowered at a velocity of $0.15 \; mm/s$ through a distance of 1.5 times the bubble radius from its initial position and held at that final position. A camera at the top (IDS UI-3060CPC) recorded the thin film interferograms (Fig.\ \ref{fig:ExperimentalSetup}c), and was illuminated with a white light source (CCS Inc. LAV-80SW2) that passed through a dichroic filter (Edmond Optics \#87245, peak pass bands: 457nm, 530nm and 628nm). The film thickness was obtained by mapping the colors in the recorded interferograms to physical thickness using the classical light intensity - film thickness relations \cite{sheludko1967thin} assuming homogeneous and non dispersive films. A Python $2.7$ based software was developed in-house \cite{frostad2016dynamic} to aid thickness mapping and visualizing the thickness profiles.

\section{Results and Discussion}\label{sec:results}

\subsection{Rheology of the WLM solutions}\label{sec:rheology}

The dynamic oscillatory rheology measurements of the tested WLM solutions over a frequency range of $6.28 \times 10^{-3}$ to $6.28$ rad/s is shown in Fig.\ \ref{fig:rheology}a and Fig.\ \ref{fig:rheology}b. The tested WLM solutions were selected from literature \cite{shikata1987micelle, inoue2005nonlinear, wheeler1996structure} so as to have the elastic ($G'$) and viscous modulus ($G''$) of the samples span about three orders of magnitude. The cross-over frequency, which is equivalent to the relaxation time for these simple Maxwell fluids, spans about two orders of the magnitude. For ease of observation, the tested samples have been displayed in two sub-figures, with Fig.\ \ref{fig:rheology}a showing samples having a low surfactant concentration ($C_D<0.01\; M$), while Fig.\ \ref{fig:rheology}b shows samples having a high surfactant concentration ($C_D = 0.1\; M$). 

The dynamic viscosity of the tested samples, as a function of shear rate over a range of $4 \times 10^{-3}$ to $10^{3}$  Hz, is shown in Fig.\ \ref{fig:rheology}c. As expected the samples have a zero shear viscosity that changed non-linearly with the salt and surfactant concentration \cite{candau1993rheological}. Above a critical shear rate all the samples were shear thinning. 

\subsection{Drainage of thin films}\label{sec:drainage}
The rupture dynamics (death) of bubbles in worm-like micellar solutions are known to deviate from those predicted by the classical models such as the Taylor-Culick model. This is a consequence of the finite bulk elasticity of WLM solutions that enhances the driving force for bubble rupture \cite{tammaro2018elasticity,sabadini2014elasticity}. In the subsequent sections we will detail the influence of bulk viscoelasticity on the early life of bubbles in WLM solutions.

As a bubble approaches the air-solution interface, a thin liquid film of thickness $h(r,\theta,t)$ separates the bubble from the ambient air (Fig.\ \ref{fig:ExperimentalSetup}b). Depending on the properties of the draining fluid and the interfaces sandwiching this fluid film, a thick region referred to as the dimple \cite{joye1992dimple,frankel1962dimpling} can be formed in the film (Fig.\ \ref{fig:ExperimentalSetup}c). The life time of the bubble is dictated by the spatiotemporal evolution (drainage) of this liquid film. The vivid dynamics accompanying the thin film drainage of the tested worm-like micelles can be seen in the accompanying \bit{Supplementary Video V1}. Interestingly, we observed a number of intriguing features in the drainage of WLM solutions. For WLM systems with very low elastic modulus, the drainage was similar to that of pure surfactant solutions. However, as the elastic modulus increased, we observed features including dimple recoil, thin film ripping, streaming as well as single single step thinning. These observations are discussed in more detail in the subsequent sections, and the initial drainage is quantified first.   

\begin{figure*}[!th]
\centering
\includegraphics[width=\linewidth]{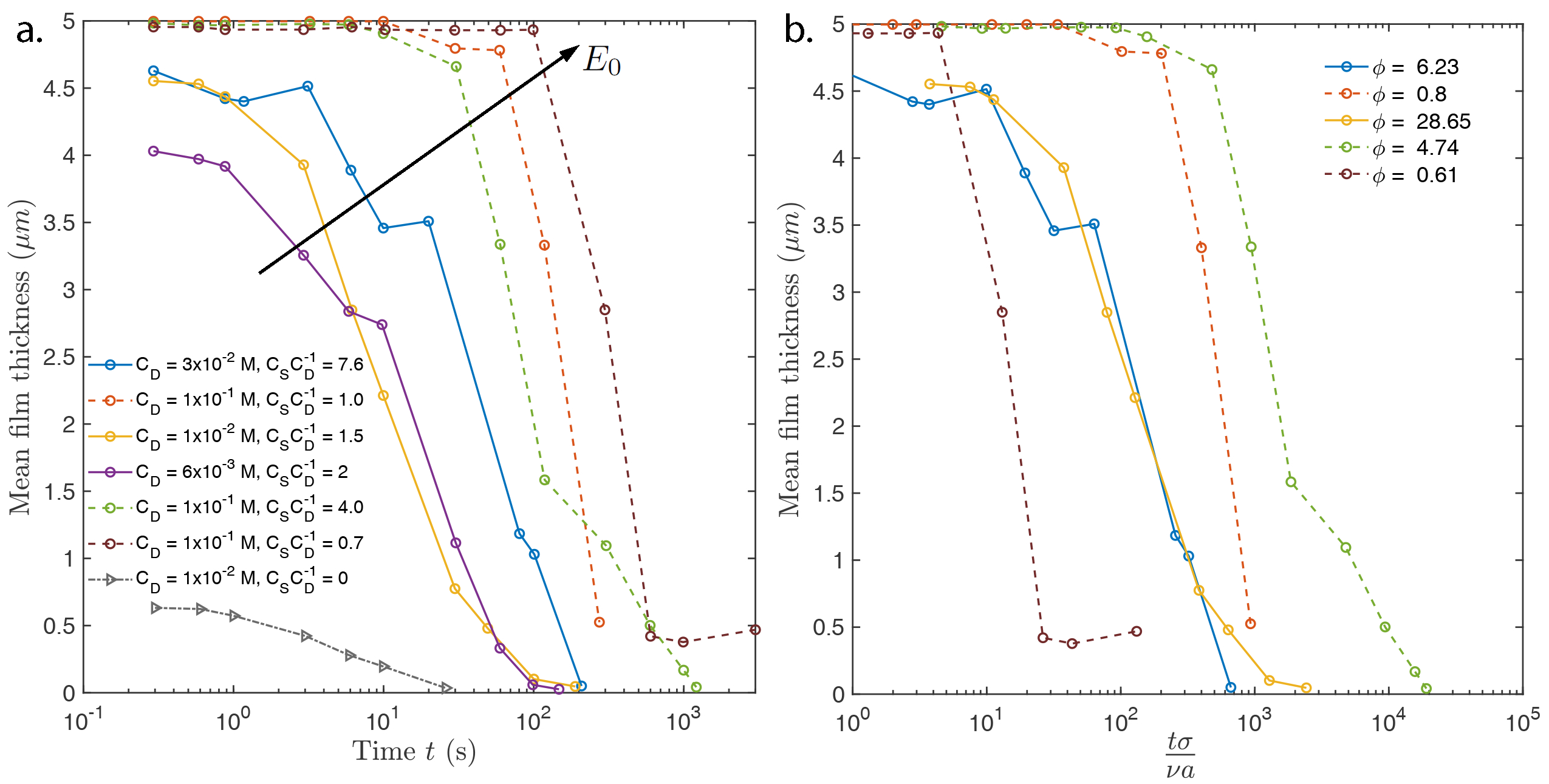}
\caption{Drainage of wormlike micelles. {\bf a.} Evolution of the mean film thickness over time for the different tested wormlike micellar solution and a pure surfactant solution. The solid lines correspond to samples in Fig.\ \ref{fig:rheology}a, the  dashed lines correspond to samples in Fig.\ \ref{fig:rheology}b and the dash-dot line corresponds to the pure surfactants. $E_0$ is the plateau modulus of the wormlike micellar solutions. {\bf b.} Mean film thickness evolution as a function of time non-dimensionalized by the capillary time scale $\tau_c = \frac{\sigma}{\nu R}$. Here $\sigma$ is the surface tension, $\nu$ is the zero shear viscosity and $R$ is the radius of the bubble. The legend indicates the values of $\phi$ for the system, where $\phi = \lambda/\tau_c $ is physically identical to the Deborah number of the system, and is calculated as the relaxation time ($\lambda$) of the WLM solutions non-dimensionalized by the capillary time scale. Note that the saturation of the obtained mean film thickness at a value of $5\;\mu m$ is a result of the limited thickness resolution range ($<5\;\mu m$) of the employed interferometric technique.}
\label{fig:drainage}
\end{figure*}

A convenient way to quantify the initial drainage dynamics is through computing the mean film thickness $\bar{h}(t)  = (\pi R_0^2)^{-1} \iint h(r,\theta,t) rdrd\theta$ \cite{suja2018evaporation,suja2020foam,kannan2019linking}. In Fig.\ \ref{fig:drainage}a, we report $\bar{h}(t)$ for the different tested WLM solutions.  From Fig.\ \ref{fig:drainage}a, we clearly observe that thin films of WLM solutions drain very differently from those comprised of pure surfactants.  Firstly, the initial mean film thickness after the bubble comes to a rest at the air-solution interface was close to an order of magnitude higher in the WLM solutions. Note that the saturation of the obtained mean film thickness at a value of $5\;\mu m$ is a result of the limited thickness resolution range ($<5\;\mu m$) of the employed interferometric technique.  Secondly, the WLM solutions drained three to four orders of magnitude slower than pure surfactant solutions. Among the WLM solutions, we also observe that the mean drainage rate is inversely correlated to the peak elastic modulus $E_0$ (see \bit{Supplementary Information Table} for the numerical values).

To further characterize the elastic effects on thin film drainage, it is necessary to discount the viscous effects on drainage. As the drainage in the initial stages is completely mediated by capillary forces, a convenient way to discount the effect of the viscous forces is to normalize the drainage time by the capillary time scale $\tau_c = \frac{\sigma}{\nu R}$. Here $\sigma$ is the surface tension, $\nu$ is the zero shear viscosity and $R$ is the radius of the bubble. The evolution of the mean film thickness as a function of the non-dimensional capillary time is plotted in Fig.\ \ref{fig:drainage}b. Note that we have not plotted the WLM sample with $C_D = 0.1 M, C_SC_D^{-1} = 2$ as it neither has a well-defined zero shear modulus nor a well-defined relaxation time in the measured frequency range. The differences in the drainage curves in the capillary scale are due to elastic effects. To build intuition into the results in Fig.\ \ref{fig:drainage}b and to appropriately quantify the elastic effects, it is helpful to consider a model system for thin film drainage. Assuming a flat film model with free slip at the boundaries and a upper convected Maxwell model for the WLM solution, the non-dimensional thickness of the film can be shown to satisfy the following governing equation with  a single parameter $\phi$ \cite{bousfield1989thinning},     
\begin{equation}\label{eq:thinFilmDrainage}
     2\phi\hat{h}^2\frac{d^2\hat{h}}{d \hat{t}^2} +  4\phi^2\left(\frac{d\hat{h}}{d \hat{t}}\right)^3 - 3\hat{h} \left(\frac{d\hat{h}}{d \hat{t}}\right)^2  - 2\hat{h}^2 \frac{d\hat{h}}{d \hat{t}}  =  0.
\end{equation}
Here $\hat{h}$ is the thickness of the film non-dimensionalized by the initial height of the film, $\hat{t}$ is time non-dimensionalized by the capillary time scale $\tau_c = \frac{\sigma}{\nu R}$ and $\phi$ is physically identical to the Deborah number and is relaxation time of the WLM solutions non-dimensionalized by the capillary time scale. In the limit of $\phi \rightarrow 0$, we recover the Newtonian result. Note that the above equation does not account for the disjoining forces and hence is only valid when the film thicker than $100$ nanometers \cite{hahn1985effects}. See \bit{Supplementary Materials} for the derivation of Eq.\ref{eq:thinFilmDrainage}.

\label{subsubsub:Recoil}
\begin{figure*}[!th]
\centering
\includegraphics[width=\linewidth]{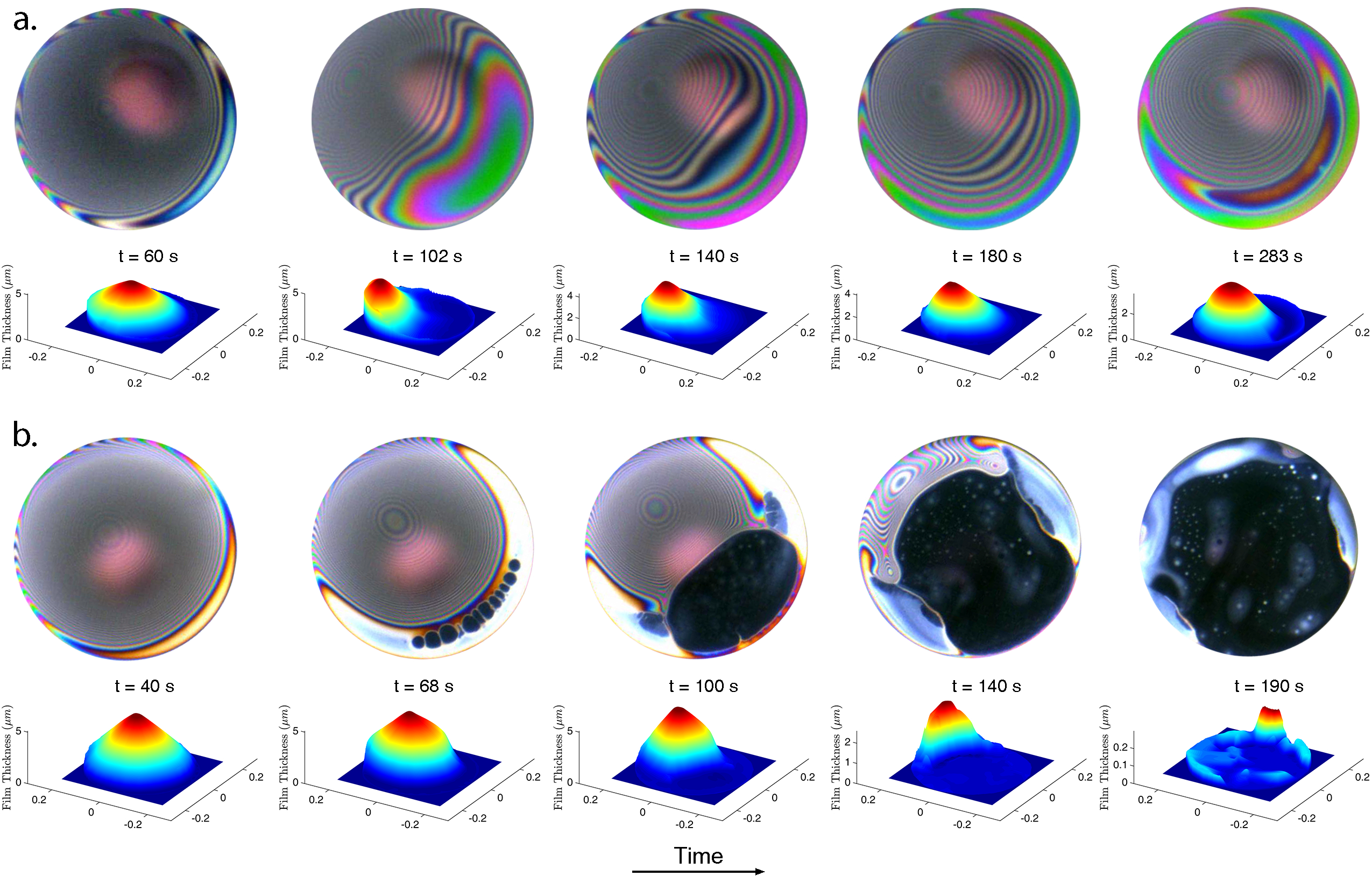}
\caption{A sequence of interferograms observed over a bubble in a worm like micellar  solution with $C_D = 0.1 M, C_SC_D^{-1} = 4$ and showing the elastic recoil of the dimple. The corresponding reconstructed physical film thickness profiles and time stamps  are indicated below the interferograms with $t=0$ indicating the time at which the bubble encounters the flat air-water interface. Note that the saturation of the obtained mean film thickness at a value of $5\;\mu m$ is a result of the limited thickness resolution range ($<5\;\mu m$) of the employed interferometric technique.}
\label{fig:elasticrecoil}
\end{figure*}

Physically, as $\phi$ increases, the contribution for the elastic stresses increase and the drainage rate is expected to fall (See \bit{Supplementary Materials} for solution of  Eq.\ref{eq:thinFilmDrainage} for various values of $\phi$). Indeed, we observed this to be the case for WLM solutions for the moderate values of $\phi$ (WLM samples denoted by the dashed lines). At small values of $\phi$ the drainage characteristics are similar to those of viscous fluids. However, as $\phi$ increases we observed fascinating elastic effects such as dimple recoil (see Section \ref{subsubsub:Recoil}). At higher values for $\phi$ (WLM samples denoted by the solid lines), we observed that the drainage rate starts to increase. This is expected as the worm-like micellar solutions can dramatically shear thin at high stresses. In addition to this expected effect, a closer look into the videos collected from the top camera (see \bit{Supplementary Video V1}) reveals that in these cases where the capillary stresses are many times larger than the elastic modulus of the WLM solution, the WLM films also fail structurally below a critical thickness. The resulting dynamics resemble the ripping of a fabric and hence will be referred to as `ripping' in the subsequent sections in the manuscript.  This structural failure along with the shear thinning nature of the WLM solutions are a possible explanation of the increased drainage we observe in samples with very high values of $\phi$.

\subsubsection{Elastic effects in thin film flows}
As dimple recoil and thin film ripping are features that have not been reported before, we will briefly discuss them further in this section. 

In Fig.\ \ref{fig:elasticrecoil}a we show a time sequence detailing the phenomenon of dimple recoil. Initially, similar to a surfactant solution, the dimple washes out to the periphery. However, the elastic effects prevent a complete washout. As the film thins and the capillary stresses decrease, the dimple retracts back to the center of the bubble. Note this elastic recoil is dramatically different from the known phenomenon of spontaneous dimpling \cite{suja2018evaporation,velev1993spontaneous,shi2020oscillatory}. Unlike spontaneous dimpling where the dimple repeatedly washes out and reforms, during the phenomenon of elastic recoil, the dimple partially washes out and retracts to the bubble apex once during the life time of the bubble.

In Fig.\ \ref{fig:elasticrecoil}b we show a time sequence detailing the phenomenon of thin film ripping. This phenomenon, similar to dimple recoil, starts with the partial washout of the dimple. Unlike the latter, as the capillary forces far exceed the elastic modulus of the WLM (indicated by the large values of $\phi$), the dimple does not recoil. Instead, below a critical thickness, the film structurally fails near the thinnest region of the film. As the ripping of the thin film progresses, ejected micellar chunks (revealed as white spots) are observed to stream around the thin film (see \bit{Supplementary Video V1}). The ripping process is completed when the dimple is completely washed out, leaving behind a black film. A consequence of this ripping process is that the WLM film progresses to black film in a single step-wise manner. This curious feature of WLM solutions observed in the terminal phases of drainage is discussed in the next section.     

\subsection{Step-wise thinning and domain expansion dynamics}\label{sec:rules_submission}

\begin{figure}[!h]
\centering
\includegraphics[width=\linewidth]{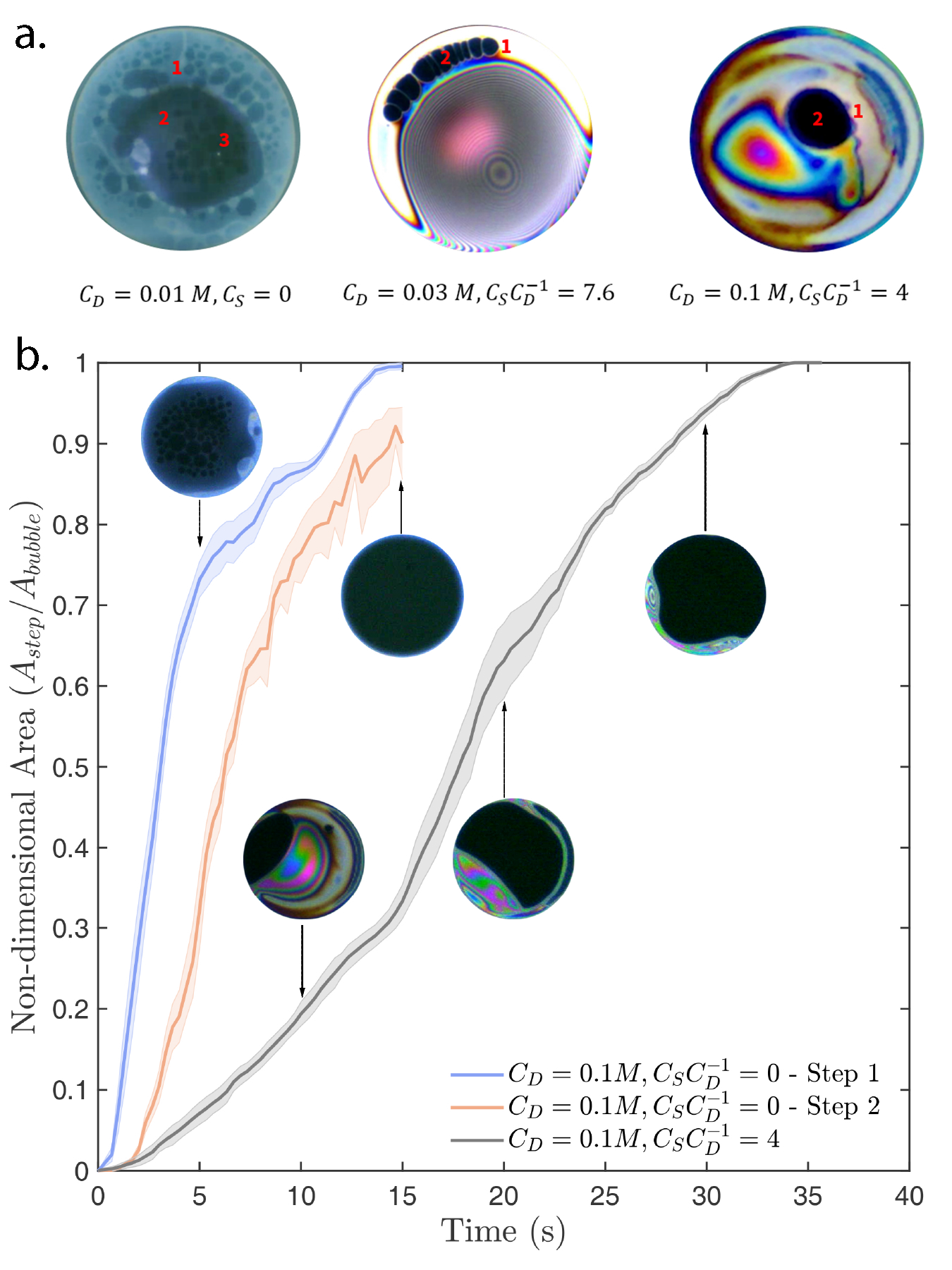}
\caption{Single step thinning in wormlike micellar solutions. {\bf a.} Terminal stages of drainage showing step-wise thinning of the thin fluid film for a pure CTAB solution ($C_D=0.01M$) and WLM solutions with  $C_D=0.03M$, $C_SC_D^{-1}=7.6$ and $C_D=0.1M$, $C_SC_D^{-1}=4$. The numbers denote the steps of decreasing film thicknesses. {\bf b.} The spatiotemporal expansion of the step wise formed newton black films for bubbles in a pure surfactant ($C_D=0.01M$) and wormlike micellar solution ($C_D=0.1M$, $C_SC_D^{-1}=4$). Bubbles in the pure surfactant solution are observed to thin via multiple steps, while the bubbles in wormlike micellar solution thins via single step. The error bars are standard deviations of two non dimensional step areas obtained using binarization thresholds varying by $10\%$. See \bit{Supplementary Video V2} for a side-by-side comparison of the terminal thinning of surfactant and WLM films.}
\label{fig:stepwisethinning}
\end{figure}

Fig.\ \ref{fig:stepwisethinning}a shows representative snapshots of the final stages of thinning for a pure CTAB and WLM solutions. CTAB solution, in the absence of salt, exhibited step-wise drainage with a step-size of around 10 nm, similar to what was seen for sodium dodecyl sulfate (SDS) \cite{zhang2015domain}. This step-wise hole opening and thinning is attributed to the structural disjoining pressure and breakup of layers of micelles trapped in the thin film. In the case of worm-like micelles, two significant differences were observed. First, instead of multiple smaller steps, the transition to a black film is in a single step from a thickness of around 80 nm to a Newton black film. Moreover, the kinetics of hole opening depended on the bulk rheological properties of the solution as described in the earlier sections. This is further elucidated in Fig.\ \ref{fig:stepwisethinning}b, which shows the spatial expansion of the black film, with time, following micellar breakage. The pure surfactant solution drains in multiple steps, and two of the steps are tracked here. Whereas the thin films of WLM fluids drain in a single step. It can be seen in Fig.\ \ref{fig:stepwisethinning}b that the kinetics of hole expansion is slower for the WLM as a result of the increased viscoelasticity. Yilixiati et al. have investigated the effect of salt on the step-wise thinning of ionic surfactant (SDS) solutions \cite{yilixiati2018influence}. They also observed a decrease in the number of steps on the addition of salt to the surfactant solutions, due to differences in the supramolecular oscillatory disjoining pressures induced by the salt. However, the step size, in their case, decreased with salt concentration. In our case, the step size is representative of the supramolecular structure of the worm-like micelle and therefore is higher than the pure CTAB micelle case. This step size of around 50 nm is identical to the mean lengths of the CTAB/NaSal worm-like micelles, studied at a variety of $C_SC_D^{-1}$  values using small angle neutron scattering \cite{aswal1998small, das2012shape, goyal1991shapes}. Similar jumps in film thicknesses have been observed for other WLM fluids  \cite{klitzing2001comparison}. Therefore, for simple surfactant solutions, the step size is lower and roughly corresponded to the size of the surfactant micelle, whereas in the case of WLM, this higher value corresponded to the length scales observed in these worm-like living polymers.

\section{Conclusions}\label{sec:conclusion}
In summary, we have reported the thin film drainage dynamics leading up to the bubble coalescence in wormlike micellar (WLM) solutions. The results revealed a number of striking characteristics of film drainage in WLM solutions that are dramatically different from those of bubbles in pure surfactants. Firstly, as expected, the drainage rate of thin films of WLM solutions are dramatically slower than those of pure surfactants due to the high bulk viscoelasticity. As a result, bubbles in WLM solutions are more stable against coalescence, with their lifetimes approximately scaling with the inverse of the peak elastic modulus. Secondly, the supramolecular structure and elasticity of WLM solutions alters the common sequence of events associated with thin film \cite{kannan2018monoclonal}. Notably, we observed a phenomenon we call as dimple recoil, whereby the partially washed out dimple is elastically pulled back to the bubble apex. In certain WLM solutions, beyond a critical thickness, we also observed a structural failure of the film that we termed as thin film ripping. Finally, the terminal drainage characteristics of WLM solutions are also different. Interestingly, WLM films shows single step thinning with step sizes comparable to the characteristic sizes of the WLM. The expansion dynamics of the Newton black domains are also different with the elastic effects dramatically altering the spatiotemporal evolution of the newton black spots.   

The results from the current study has two important implications for scientific efforts focused on understanding of life and death of bubbles \cite{debregeas1998life}. Firstly, we have revealed the fascinating dynamics at play on the surface of bubbles in WLM solutions during its life time. These results complement prior research that has documented the interesting events that accompany the death (rupture) of bubbles in WLM solutions \cite{tammaro2018elasticity,sabadini2014elasticity}, and thus gets us closer to completely understanding the life and death of bubbles in WLM solutions. These results also supplement the significant work that has been done in the past on the dynamics and coalescence of bubbles in viscoelastic media \cite{gaudron2015bubble, acharya1978note, dekee1986bubble,klitzing2001comparison,leal1971motion}. Secondly, our results reveal a hitherto undocumented single step-wise thinning phenomenon in WLM solutions. This observation complements and opens up new venues for the decades long research looking at the terminal drainage characteristics of thin films \cite{zhang2015domain,yilixiati2018influence,kralchevski1990formation}.   

There remains several opportunities for future work that would
offer important extensions to the present work. First and foremost, future studies could address limitation of the current study in probing the drainage dynamics of films thicker than $5\; \mu m$. Computational or experimental studies may be designed to study drainage dynamics WLM films larger than $5\; \mu m$. Secondly, future studies could systematically study the observed single step thinning phenomenon as function of the surfactant and salt concentration, and reveal the mechanisms at play.

\bibliographystyle{vancouver}
\bibliography{References}

\end{document}